\documentclass[12pt]{article}      
\usepackage{amsmath,amssymb} 
\usepackage{color}                    
\usepackage{hyperref}      

\numberwithin{equation}{section} 

\newcommand{\half}{\frac{1}{2}}
\newcommand{\be}{\begin{equation}}
\newcommand{\ee}{\end{equation}} 
\newcommand{\ba}{\begin{eqnarray}}
\newcommand{\ea}{\end{eqnarray}}
\newcommand{\nn}{\nonumber}
\newcommand{\N}{{\cal{N}}}
\newcommand{\Tr}{{\rm{Tr}}}
\newcommand{\del}{\partial}
\newcommand{\delb}{\bar{\partial}}
\newcommand{\Zb}{\mathbb{Z}}
\newcommand{\Cb}{\mathbb{C}}

\begin{document}


\thispagestyle{empty}

 \renewcommand{\thefootnote}{\fnsymbol{footnote}}
\begin{flushright}
 \begin{tabular}{l}
 {\tt arXiv:0904.1460[hep-th]}\\
 {SNUTP 09-007}
 \end{tabular}
\end{flushright}

 \vfill
 \begin{center}
 {\bfseries \Large Surface operators in the Klebanov-Witten theory}
\vskip 1.9 truecm

\noindent{{\large
  Eunkyung Koh \footnote{ekoh(at)phya.snu.ac.kr}
and Satoshi Yamaguchi \footnote{yamaguch(at)phya.snu.ac.kr} }}
\bigskip
 \vskip .9 truecm
\centerline{\it Department of Physics and Astronomy,
Seoul National University,
Seoul 151-747, KOREA}
\vskip .4 truecm
\end{center}
 \vfill
\vskip 0.5 truecm

\begin{abstract}
We investigate 1/2 BPS conformal surface operators in the Klebanov-Witten theory. 
These surface operators preserve certain parts of the conformal symmetry and R-symmetry as well as half of the supersymmetry. We propose the gravity dual of the surface operator as a configuration of a D3-brane in $AdS_5\times T^{1,1}$. This D3-brane preserves the same amount of the supersymmetry as the surface operator.  We also compute the correlation function of the surface operator and a chiral primary operator.
\end{abstract}

\vfill
\vskip 0.5 truecm

\setcounter{footnote}{0}
\renewcommand{\thefootnote}{\arabic{footnote}}

\newpage

\tableofcontents

\section{Introduction}

A surface operator is an operator supported at a surface defined by a condimension 2 singularity. Disorder type surface operators in $ \N=4$ SYM have been introduced in \cite{Gukov:2006jk}.  In \cite{Gukov:2006jk}, the gravity dual of the operators have been proposed  as probe D3-branes wrapping $ AdS_3 \times S^1$ in  $ AdS_5 \times S^5$, which can be supersymmetric \cite{Skenderis:2002vf, Constable:2002xt}.  The gravity dual of these operators are also described by the bubbling AdS geometry\cite{Lin:2004nb,Lin:2005nh,Gomis:2007fi}.  Some of the physical quantities have been evaluated in both gauge theory and gravity \cite{Drukker:2008wr}.

Along with the canonical example  $AdS_5 \times S^5$/$\N=4$ super Yang-Mills \cite{Maldacena:1997re}, the AdS/CFT duality with less supersymmetries  also has been studied, which one can refer to \cite{Aharony:1999ti} and references there in. Symmetry-wise, $AdS_5$ gravity with an internal Sasaki-Einstein  5-manifold can be dual to $\N=1, d=4$ supersymmetric CFT . 
One of the first examples is given by Klebanov and Witten \cite{Klebanov:1998hh}, for the dual background $AdS_5 \times T^{1,1}$. 

In this paper, we will present a new example of $1/2$ BPS conformal surface operators in $ \N=1$ superconformal theories, especially in the Klebanov-Witten theory. We use $F$ term and $D$ term constraints to define the surface operator. The surface operator in this paper is defined by a complex singularity with a fractional power. We will show that an appropriate gauge transformation can cancel the monodromy around the singularity, and thus the boundary condition can be well-defined. 

We will propose that the gravity dual of the operator is a probe D3-brane wrapping $AdS_3 \times S^1$ in $AdS_5 \times T^{1,1}$.  We show that the brane preserves the same symmetries with the surface operator.  

For BPS surface operators in $ \N=4$ SYM, the correlation functions of BPS surface operators with chiral primary operators, evaluated in the gravity side by using GKPW prescription  in SUGRA limit \cite{Gubser:1998bc, Witten:1998qj} , have been shown to agree with the results from the gauge theory in the semi-classical limit \cite{Drukker:2008wr}.  We will consider correlation functions of the surface operator and CPO's in Klebanov-Witten theory. The evaluation  in gravity  can be done  in the large $AdS$ radius limit,  as in the $ \N=4$ SYM case. The evaluation in the gauge theory, however,  becomes difficult, since the theory is intrinsically strongly coupled.  Instead,  we will use the results in gravity to deduce a possible form of the normalization factor of CPO's as a function of the parameters of the theory, on the assumption that the semi-classical approximation is valid in the gauge theory. 

The organization of this paper is as follows. In section \ref{sec:gauge}, we construct the $1/2$ BPS surface operator in the Klebanov-Witten theory. In section \ref{sec:gravity}, we propose the gravity dual of the surface operator as a probe D3-brane. Using kappa symmetry projection we show that a probe D3-brane wrapping the $U(1)$ isometry direction in a Sasaki-Einstein 5-manifold preserves $1/2$ of the fermionic symmetries. We evaluate correlation functions of the surface operator with chiral primary operators by using the D3-brane solution.  Some discussions will follow in section \ref{discussion}.

\section{An example of $1/2$  BPS surface operator in Klebanov-Witten theory} 
\label{sec:gauge}

\subsection{Definition of the surface operator} \label{KW} 

The Klebanov-Witten theory is a certain  
$\N=1$ supersymmetric gauge theory with gauge group  $SU(N) \times SU(N)$   at its IR fixed point, which is known to be related to D3-branes on the tip of the conifold  \cite{Klebanov:1998hh}. 
The bosonic fields in the theory are  chiral superfields  $A_i$ ($B_i$), for $i=1,2$,   in the (anti-)bi-fundamental  representation, and gauge fields $A_{ \mu} , \tilde{A}_{ \mu} $ for each $SU(N)$.  The superpotential is given as 
\be W = \Tr [ A_1 B_1 A_2 B_2 - A_1 B_2 A_2 B_1] .  \label{super_p}  \ee

 Let us consider the theory on  $ { \mathbb R}^{4} $ with coordinates $ (x^1, x^2, x^3,x^4)$, or ${ \mathbb C}^2 $ with $ (z_1, z_2)$. We define  complex coordinates $z_1= x^1+ i x^2, z_2 = x^3+ i x^4$.  Sometimes we will use polar coordinate of $z_1$-plane, $ z_1 = r e^{ i \phi } $.  For the $\N=1$ super-space, 
 unless otherwise stated, we will use the conventions of Wess and Bagger \cite{Wess:1992cp}. 

As in \cite{Gukov:2006jk}, we characterize a surface operator by the boundary condition of bosonic fields near codimension 2 singularities. In this paper, we will consider a surface operator supported at $ z_1 = 0 $.  The situation is similar to \cite{Drukker:2008jm} where operators from codimension 2 singularities of the bi-fundamental scalar fields in ABJM theory \cite{Aharony:2008ug} are considered.

The behavior of the fields at the singularity is determined in such a way that the surface operator preserves the dilatation symmetry. The conformal dimension of $A_j$ and $B_j$ is $3/4$ as can be seen in the form of the superpotential \eqref{super_p}. Thus $A_j$ and $B_j$ have the singularity of $z_1^{-3/4}$. 

More concretely, we consider the following classical configuration.
\begin{align} 
&  A_1 = B_1 =  \frac{ \beta}{ z_1^{ 3/4} } diag. ( 1, i, 0, \cdots , 0)  ,\qquad A_2=B_2=0,\qquad (\text{gauge fields})=0, \label{example}
\end{align} 
where $\beta$ is a real constant parameter. This is the main example of the surface operator considered in this paper. 

The configuration \eqref{example} is not single-valued under $z_1\to z_1e^{2\pi i}$ rotation. Since this monodromy is a part of the gauge transformation, it is canceled by introducing the gauge holonomy $(g, \tilde{g}) \in SU(N)\times SU(N)$ around $z_1=0$ of the following form.
\begin{equation}
\begin{aligned} 
&  g =\left(
\begin{array}{cc}
 \sigma_1 & 0 \\
 0 & e^{\frac{\pi i}{N-2}}I_{N-2} 
\end{array}
\right),\qquad
  \tilde{g} =\left(
\begin{array}{cc}
 i\sigma_2 & 0 \\
 0 & I_{N-2} 
\end{array}
\right),\\
&\sigma_1:=\left(\begin{array}{cc}
0 &1 \\
1 &0 \\
\end{array}\right),\qquad
i\sigma_2:=\left(\begin{array}{cc}
0 &-1 \\
1 &0 \\
\end{array}\right),\qquad
I_{N-2}: ( (N-2)\times (N-2)\text{identity matrix}).
\end{aligned}
\nn 
\end{equation}
 $A_j$ and $B_j$ are transformed as
\begin{align}
 A_j\to g A_j \tilde{g}^{-1},\qquad
 B_j\to \tilde{g} B_j g^{-1}.\qquad
\end{align}
This introduction of the holonomy is the similar trick as \cite{Koh:2008kt}.

There are rather trivial generalizations of this surface operator \eqref{example}. One is the phase of $\beta$. The other is to introduce the gauge field
\begin{align}
 A=\tilde{A}= d\phi
\left(
\begin{array}{cc}
 \alpha I_2 & 0 \\
 0 & -\frac{2\alpha}{N-2}I_{N-2} 
\end{array}
\right),
\end{align}
where $\alpha$ is a real constant. We omit these generalization parameters in the rest of this paper.

It is possible to consider BPS surface operators which do not preserve the dilatation symmetry. In that case, one can choose different order of the singularity of $z_1$, analogously to the surface operators in $\N=4$ SYM with higher poles \cite{Witten:2007td}. The dilatation symmetry is useful when we consider more general surface operators which depend on $z_2$ as well as $z_1$, since one can restore a time-like direction by the Wick rotation of the radial direction of ${\mathbb R}^4$ as ${ \mathbb R}_{ time} \times S^3$.  

Let us see the bosonic symmetry of the configuration \eqref{example}. This configuration is invariant under global conformal symmetry of two dimensions $(x^3,x^4)$, which is SO$(1,3)$ (or SO$(2,2)$ in Lorentzian signature). It also preserves the following $U(1)^3$. Let $J_A^3,J_B^3$ denote respectively the 3rd generators (generated by $\sigma_3/2$) of the two SU$(2)$ global symmetries: SU$(2)_A$ and SU$(2)_B$ which respectively rotate $(A_1,A_2)$ and $(B_1,B_2)$ as doublets. Also let $R$ be the U$(1)_R$ symmetry generator ($A_j, B_j$ are charge $1/2$) and $M_{12}$ be the rotation in $(x^1,x^2)$-plane. The configuration \eqref{example} preserves the following three U$(1)$ symmetries.
\begin{itemize}
 \item U$(1)_d$: generated by $(3/2)R+M_{12}$.
 \item U$(1)_a$: generated by $J^3_A+J^3_B-R$.
 \item U$(1)_v$: generated by $J^3_A-J^3_B$ accompanied with some gauge transformation.
\end{itemize}
We will see that these symmetries $SO(2,2)\times U(1)^3$ are also present in the gravity dual.

\subsection{Supersymmetry in the gauge theory}

 Consider $ \N=1$ supersymmetric gauge theory, with a Lagrangian in the form of 
\be L =  \int d^4 x \int d^4 \theta K (  \bar{ \Phi }^{ \bar{i} } e^{- V }, \Phi^i ) + \int d^4 x \left(   \left(   \frac{1}{4  } { \cal W}^{ \alpha} { \cal W}_{ \alpha} + W ( \Phi^i ) \right) \vert_{ \theta^2 } + c.c  \right) , 
\nn \ee
for chiral superfields $ \Phi^i = \varphi^i (x)  + \sqrt{2} \theta \psi^i (x) + \cdots $ and a vector superfield $V = - \theta \sigma^{ \mu} \bar{ \theta} A_{ \mu} (x) + i \theta^2 \bar{ \theta} \bar{ \lambda}(x) + \cdots $. For the Klebanov-Witten theory, $ \varphi^i$'s represent $A_1, A_2, B_1, B_2$ in the previous section. $K$ is the K\"{a}hler potential of the theory. 
 The variation of fermions are given as follows
\begin{align}
\delta \psi^i &= i \sqrt{2} \sigma^{\mu}  \epsilon D_{ \mu} \varphi^i + \sqrt{ 2} \epsilon F^i ,  \label{quark}  \\
\delta \lambda &= \sigma^{ \mu \nu } \epsilon F_{ \mu \nu } + i \epsilon D,  \label{gaugino} 
\end{align}
while the equation of motions of $F$  and $D$ terms for non-trivial K\"{a}hler potential are as follows
\begin{align} F^i &  = \half  \Gamma^{i}_{ jk} \psi^j \psi^{k} -  g^{ i \bar{l} } \del_{ \bar{l} } \bar{W},   \label{fterm} \\
D^m & =  \Tr [ g_{ i \bar{j} }\bar{ \varphi}^{ \bar{j} } T^m_{ R( \varphi^i) } \varphi^i ],   \label{dterm} 
\end{align} 
where $g_{ i \bar{j} } \equiv  \frac{ \del}{\del \Phi^i}  \frac{ \delb}{ \delb\bar{ \Phi}^{ \bar{j} }  } K  $,  $ \Gamma^i_{ j k } = g^{ i  \bar{j} } \del_j g_{ k \bar{j} } $.  $R( \varphi^i)$ denotes the representation of $ \varphi^i$.

For our definition of a surface operator \eqref{example}, $F$ and $D$ term vanish. 
$F$ term vanishes since $\del_{ \bar{i} } \bar{ W}$ vanishes for each $\bar{i}$ and we set the boundary conditions of fermion fields vanish. The $D$ term condition \eqref{dterm} for the first gauge group $SU(N)$ can be rewritten as 
\be D^m =   \Tr [ g_{ 1 \bar{1} }^A \bar{A}^{ \bar{1} } T^m_N A^1 ] +   \Tr [g_{ 1 \bar{1} }^B \bar{B}^{ \bar{1} } T^m_{ \bar{ N } } B^1 ]   . \label{dterm:2} \ee
Note that  $T_{ \bar{N} } = - T^{ \ast}_N $ for hermitian generators and   $g_{ i \bar{j} }^A \equiv \frac{ \del}{ \del A^i} \frac{ \delb}{ \delb \bar{A}^{ \bar{j} } } K, g_{ i \bar{j} }^B \equiv \frac{ \del}{ \del B^i} \frac{ \delb}{ \delb \bar{B}^{ \bar{j} } } K $. On the assumption that the K\"{a}hler potential has the symmetry of exchanging $A_i$ and $B_i$, eq.~\eqref{dterm:2} vanishes in our configuration \eqref{example}.

We are considering a flat gauge connection,  $ F_{ \mu \nu} = 0$, thus the variation of gaugino \eqref{gaugino} vanishes.  The variation of quarks \eqref{quark} can be written as 
\be \delta \psi^i = i \sqrt{2} ( \sigma^1 + i \sigma^2 ) \epsilon \frac{   \del }{ \del {z_1} }  \varphi^i (z_1), \nn \ee
since the boundary condition for the chiral fields are holomorphic functions of $z_1$. 
Now impose the following condition on the supersymmetry parameter $ \epsilon $, %
\be ( \sigma^1 + i \sigma^2 ) \epsilon = 0 , \label{half} \ee
then  the supersymmetry variation of the quark \eqref{quark} vanishes. Thus our definition of the surface operator in \eqref{example} preserves half of the supersymmetries.

\subsection{Correlation functions with chiral primary operators}

In the semi-classical approximation, the correlation function of a local operator ${ \cal O} ( \zeta) $ and a surface operator is evaluated by taking the value in the configuration \eqref{example}.
\be \frac{ \langle { \cal O}_{ \Sigma} \cdot { \cal O } ( \zeta )  \rangle }{ \langle { \cal O}_{ \Sigma} \rangle }  \simeq { \cal O} \vert_{ \Sigma} ( \zeta ) . \label{semi} \ee

Chiral primary operators in KW theory has been discussed in \cite{Klebanov:1998hh, Herzog:2002ih}. The operators are in the following form, 
\be { \cal O}_{ \Delta}^I  = p^{I}_{\Delta}C_{ n }^{  I ( i_1, \cdots, i_{ n } ) , ( j_1, \cdots, j_{ n } ) } \Tr [ A_{ i_1} B_{j_1 } \cdots A_{ i_{ n}  } B_{i _{n}   } ]  \label{cpo} \ee
where $( i_1, \cdots, i_i)$ means symmetrized indices. $C_n^I$ is the coefficient in the corresponding spherical harmonics of $T^{1,1}$. $p^{I}_{\Delta}$ is a normalization constant to normalize the two point function of operators as
\begin{equation}
 \langle\overline{ \cal O}_{ \Delta}^I(x){ \cal O}_{ \Delta}^I(0)\rangle=\frac{1}{|x|^{2\Delta}},\label{2pt}
\end{equation}
where $\overline{ \cal O}_{ \Delta}^I(x)$ is the hermitian conjugate of ${ \cal O}_{ \Delta}^I(x)$.
The conformal dimension $ \Delta$ of ${ \cal O}^{I}_{ \Delta } $ is
\be
\Delta =  \frac{3}{2}  n , \qquad  ( \mbox{for   }n \in { \mathbb Z}_{+})  . 
\label{cdim} \ee

If we consider correlation function of CPO's with a surface operators defined by \eqref{example}, it is non-trivial only when the CPO's are in the following form,  
\be { \cal O}_{ \Delta    }  = p_{\Delta} C_{ n }  \Tr [ ( A_1 B_1)^{ n} ] ,     \ee
where $C_{n}$ is the coefficient of spherical harmonics in \eqref{sph_h}. Using \eqref{semi}, the correlation function is given as follows, 
\be  \frac{ \langle { \cal O}_{ \Sigma} \cdot { \cal O }_{ \Delta}    \rangle }{ \langle { \cal O}_{ \Sigma} \rangle }  \simeq  p_{\Delta}   C_{ n  }     \frac{   \beta^{2n} }{ ( z_1 )^{ \frac{3}{2} n }    }    \left( 1 + (-1)^n  \right) .
\label{gauge}  \ee

Actually this classical approximation is not justified very well since the theory is on the non-trivial fixed point and the quantum effect is supposed to be large. However, 
the spacetime dependences of the result can be determined by symmetry. Later we will compare the result with the result from supergravity.

\section{Gravity dual of the $1/2$ BPS surface operator} 
\label{sec:gravity}

\subsection{Probe D3-brane wrapped on a holomorphic hypersurface in $ AdS_5 \times T^{1,1} $}  \label{d3}
The background of the gravity dual of the KW theory is known as $AdS_5 \times T^{1,1}$ \cite{Klebanov:1998hh}. 
The metric of this background is \cite{Romans:1984an}
\be ds^2_{10} =     \frac{1}{y^2} (  \sum_{ A= 1,2} \vert dz_A \vert^2 +dy^2 )+ ds^2_{ T^{1,1}},  \label{10d} \ee
where the metric of the $T^{1,1}$ is given by\cite{Candelas:1989js}
\be ds^2_{ T^{1,1} } = \frac{1}{9} ( d \psi + \cos \theta_1 d \nu_1 + \cos \theta_2 d \nu_2)^2 +  \frac{1}{6} \sum_{ i = 1,2} ( d \theta_i^2 + \sin^2 \theta_i d \nu_i^2 )  , \label{metric:t11} \ee
for $ 0 \leq \psi \leq 4 \pi, 0 \leq \theta_i \leq \pi, 0 \leq \nu_i \leq 2 \pi $. In this paper, we choose the unit of length as
\begin{equation}
 (\text{AdS radius})^4=\frac{27}{4}\pi \alpha'^2 g_s N=1.\label{unit}
\end{equation}
We propose that the gravity dual of the surface operator in eq.\eqref{example} is a probe D3-brane wrapping a surface in $AdS_5\times T^{1,1}$ expressed by
\begin{align}
 &  y (r) = \frac{r}{ \kappa }, \quad \psi ( \phi)  = - 3 \phi \mod{2\pi} ,  \quad \theta_i = \pi, \quad \nu_i = 0. \quad ( i = 1, 2) \label{transverse}  
\end{align}
where $ z_1  \equiv r e^{ i \phi}$ and $\kappa$ is a constant related to $\beta$.

Let us explain the reason for this identification.
The metric cone of $T^{1,1}$ is conifold, namely 
\be ds^2_{\text{conifold}} = d \rho^2 + \rho^2 ds^2_{ T^{1,1} } .  \label{cone} \ee
On the other hand, the conifold  can be expressed as a hypersurface in $\Cb^4$ with coordinates $ ( \omega^1, \omega^2, \omega^3, \omega^4)$ defined by the equation
\be \sum_{ A=1}^4 ( \omega^A)^2 = 0 . \label{conifold} \ee
The parameterization of $ \omega^A $ in terms of $ \psi, \theta_i, \nu_i$ is given in \eqref{parameter}. 
In \eqref{cone}, $ \rho^2$ is related with $ \omega^A$ by 
$ 
 \rho^2 = y^{-2} = ( \sum_{A=1}^4 \vert \omega^A \vert^2 )^{2/3}$ \cite{Candelas:1989js}. 
Eq.~\eqref{conifold} can be solved by the following parameterization   \cite{Klebanov:1998hh}
\begin{align}
(  \omega^1 +  i \omega^4)= A_1 B_1 , \quad  ( \omega^1 - i \omega^4 ) = A_2 B_2, \quad     (  i \omega^2 + \omega^3 )  = A_1 B_2 , \quad (  i \omega^2 - \omega^3 ) = A_2 B_1  \label{parameter_2}. 
\end{align}
These coordinates $A_1,A_2,B_1,B_2$ are related to the scalar fields in the field theory. The field theory configuration \eqref{example} is $A_1=B_1\sim$(constant)$z_1^{-3/4}$. Eq.~\eqref{transverse} is this relation expressed by the coordinates in \eqref{10d}-\eqref{metric:t11}. Thus it is appropriate to identify the D3-brane \eqref{transverse} as the gravity counterpart of the surface operator \eqref{example}.

Notice that the induced metric of the D3-brane \eqref{transverse} is $AdS_3\times S^1$. Actually this configuration preserves the SO$(2,2)\times$ U$(1)^3$ symmetry discussed in section \ref{KW}. Here SO$(2,2)$ is a part of the isometry of the $AdS_5$ and preserved the same way as the case of the 1/2 BPS surface operator of ${\cal N}=4$ super Yang-Mills theory\cite{Constable:2002xt,Gukov:2006jk,Gomis:2007fi}. Three $U(1)$'s are generated by Killing vectors
\begin{equation}
 -3\frac{\del}{\del \psi}+\frac{\del}{\del \phi},\qquad
 \frac{\del}{\del \nu_1}+\frac{\del}{\del \nu_2}+2\frac{\del}{\del \psi},\qquad
\frac{\del}{\del \nu_1}-\frac{\del}{\del \nu_2}.
\end{equation}
These three Killing vectors correspond to U$(1)_d$, U$(1)_a$ and U$(1)_v$, respectively.

\subsection{Supersymmetry of the probe D3-brane}
\label{susy_brane}

In this subsection, we will check the supersymmetry of the D3-brane configuration. Actually it is rather straightforward, at least locally, to generalize this configuration to more general cases. Therefore in this subsection, we work in the general $AdS_5 \times SE_5$ background, where $SE_5$ is a Sasaki-Einstein 5-manifold.

The 10-dimensional metric is given as 
\begin{align} ds^2 & = \frac{1}{ y^2 }  ds^2_{ { \mathbb R}^{1,3} } + y^2 ds^2_{ CY_3}   = ds^2_{ AdS_5} + ds^2_{ SE_5} , \nn \end{align}
while $CY_3$ is a toric Calabi-Yau 3 fold which is the cone of $SE_5$, 
\be ds^2_{CY_3} = \frac{  d y^2}{ y^4} + \frac{1}{y^2}  ds^2_{SE_5} . \nn \ee
 The 5-form field strength in the background is given as 
\be F_5 = 4 ( vol ( AdS_5) + vol( SE_5 ) ) . 
\nn \ee
The gravitino variation in this background becomes 
\be \delta \psi_A = \nabla_{A} \epsilon + \frac{i}{4} \left( \Gamma^{01234 } + \Gamma^{56789} \right) \Gamma_A \epsilon .  
\label{gravitino} \ee
If $ \delta \psi_A =0$ for some $ \epsilon$, $ \epsilon$ is the Killing spinor of the given background.

To check the preserved supersymmetry of the probe D3-brane,
we can use the kappa symmetry projection  \cite{Cederwall:1996pv,Aganagic:1996pe,Cederwall:1996ri,Bergshoeff:1996tu,Aganagic:1996nn,Bergshoeff:1997kr}. 
The 
projection operator is given by 
\begin{align}
 \Gamma =& \frac{1}{ \sqrt{ - \det G} } \Gamma_{A_0 A_1 A_2 A_3 } E^{A_0}_{M_0} E^{A_1}_{M_1}   E^{A_2}_{M_2}  E^{A_3}_{M_3} \frac{ \partial X^{M_0} }{ \partial \xi^0 }  \frac{ \partial X^{M_1} }{ \partial \xi^1 }  \frac{ \partial X^{M_2} }{ \partial \xi^2 }  \frac{ \partial X^{M_3} }{ \partial \xi^3 }.   \label{projection}
\end{align} 
The number of preserved supersymmetries  is given by the number of Killing spinors $ \epsilon$  satisfying 
\be i \Gamma \epsilon = \epsilon.  \label{k:cond} \ee
We review the Killing spinor of $AdS_5 \times SE_5$ in appendix \ref{spinor}. 

The metric of a Sasaki-Einstein 5 manifold can be written, at least locally, as 
\be ds^2_{SE_5} = \left( \zeta d \psi + \eta \right)^2 + ds^2_{X_4},   \label{regular}  \ee
where the Reeb vector field can be written as $\frac{1}{\zeta}\frac{\del}{\del \psi}$ and $ \eta $ is an 1-form of 4 dimensional manifold $X_4$. $ \zeta $ is some constant.  
We define a vielbein $E^5$ of 10 dimensional space-time by 
$ E^5 = \zeta d \psi + \eta$, 
where other vielbeins are defined by $ ds^2_{ AdS_5} = \sum_{ a = 0}^4 E^a \otimes E^a$, and $ds^2_{ X_4} = \sum_{ a= 6}^9 E^a \otimes E^a$.

We choose the world-volume coordinate of the D3-brane as 
\be \xi^m = \{ x^0, x^1, x^2, x^3 \}.   \label{world} \ee
The ansatz for the transverse directions are, 
\begin{align} 
& y(r) = \frac{r}{ \kappa}  , \quad  \psi ( \phi)  = -  \frac{1}{ \zeta } \phi ,  \quad  \mbox{coordinates of } X_4 = constant.   \label{ansatz}
\end{align} 
The projection operator \eqref{projection} in our ansatz \eqref{ansatz} becomes 
\ba \Gamma 
&=& \frac{1}{ 1 + \kappa^2} \Gamma_{03} \left( \Gamma_{45} + \kappa \cos \phi  ( \Gamma_{15} + \Gamma_{24} ) + \kappa \sin \phi ( \Gamma_{25} - \Gamma_{14} ) - \kappa^2 \Gamma_{12} \right),  \nn 
\ea
where $ \Gamma_A$ are gamma matrices of the tangent space, $ \{ \Gamma_A , \Gamma_B \} = 2 \eta_{ AB} $. The condition \eqref{k:cond} can be rewritten as
\ba
\epsilon_{ \pm} 
&=& \frac{ \pm}{ 1 + \kappa^2} \left( ( \Gamma_{1245} + \kappa^2 ) \epsilon_{ \pm} + \kappa \cos \phi ( - \Gamma_{25} + \Gamma_{14} ) \epsilon_{ \mp} + \kappa \sin \phi ( \Gamma_{15} + \Gamma_{24} ) \epsilon_{ \mp}  \right)  , 
\label{eps}
\ea
where $ \epsilon_{ \pm}$ are projections of $ \epsilon$ by  $ \epsilon_{ \pm} = \half ( 1 \pm i \Gamma^{0123} ) \epsilon$. \eqref{eps} can be satisfied if the followings hold, 
\begin{align}
& \Gamma_{1245} \epsilon_{-} = - \epsilon_{-} , \nn \\
& \Gamma_{1245} \epsilon_{+} = \epsilon_{+}  - 2 \kappa ( \cos \phi \Gamma_1 + \sin \phi \Gamma_2 ) \Gamma_4 \epsilon_{-} . \nn 
\end{align}
In terms of the 4 component spinors $ \chi$  \eqref{k:se5}  and  $ \eta_{ \pm}$ \eqref{lambda},  which are in the decomposition of the Killing spinor $ \epsilon$ of $AdS_5 \times SE_5$  \eqref{decomposed}, \eqref{lambda}, the above  can be rewritten as  
\begin{align}
 i \tilde{ \Gamma}_{ 12} \eta_{-} \otimes \tilde{ \Gamma}_5 \chi &= \eta_{-} \otimes \chi  \label{eq1} \\
i \tilde{ \Gamma}_{12} \eta_{+} \otimes \tilde{ \Gamma}_5 \chi &= \eta_{+} \otimes \chi + \left( 2 ( x^1 \tilde{ \Gamma}_1 + x^2 \tilde{ \Gamma}_2  ) - x^{ \mu} \Gamma_{ \mu} \right) \left( - i \tilde{ \Gamma}_{12} \eta_{-} \otimes \tilde{ \Gamma}_5 \chi + \eta_{-} \otimes \chi \right) 
\label{eq2}
\end{align}
where $ \tilde{\Gamma}_A$ are defined in \eqref{tilde}. 
In the appendix \ref{proof},  we will prove that 
\be \tilde{ \Gamma}^5 \chi = - \chi . \label{eqn}  \ee
 We now  impose the following condition for constant spinors $ \eta_{ \pm}$, 
\be \tilde{ \Gamma}_{ 12} \eta_{ \pm} = i \eta_{ \pm}.  \nn \ee
Using \eqref{eqn}, one can show \eqref{eq1} and \eqref{eq2} hold under the imposition. 
Thus the probe D3-brane wrapping on  \eqref{ansatz} preserves the half of the fermionic symmetries. 

$Y^{p,q}$ and $L^{p,q,r}$ are other examples of Sasaki-Einstein 5-manifold with explicitly known metric,  constructed in \cite{Gauntlett:2004yd,Cvetic:2005ft,Cvetic:2005vk,Martelli:2005wy}.   Supersymmetric D3-brane probes wrapped on a 2 cycle or 3 cycle in $T^{1,1}$, $Y^{ p, q}$ and $L^{ p,q,r}$  have been studied in  \cite{Arean:2004mm, Canoura:2005uz, Canoura:2006es, Edelstein:2007bx}. 

\subsection{Correlation functions with chiral primary operators } \label{cpo:gravity} 

We now consider correlation functions of the surface operator in \eqref{example} with chiral primary operators. As in \cite{Drukker:2008wr, Koh:2008kt}, we use the GKPW prescription \cite{Gubser:1998bc, Witten:1998qj} to get the one-point function of the chiral operators . We introduce the boundary source $s_0^{\Delta}$ which induce the solution of the linearized equation of motion of the bulk scalar field $s$. The solution can be written as
\begin{align} 
& s_{\Delta}(y,x,\psi,\nu_i) =   \int d^4 x' G ( y, x: x^{ \prime}) Y_{ n   } ( \psi,  \nu_i  ) s^{ \Delta    }_0(x'), \label{source} 
\end{align} 
where $ x^{ \prime}$ is the position of source on the boundary of $AdS_5$, and $G(y,x:x^{ \prime})$ is the bulk-boundary propagator, 
\be G (y, x ; x^{ \prime} )  = c ( \Delta )  \frac{ y^{  \Delta }  }{ ( y^2 +\sum_{ i  =1}^4 ( x^{ i } - x^{ i  \prime} )^2  )^{  \Delta} },  \nn \ee
for a scalar field of which equation of motion is 
\be \nabla_{ \mu} \nabla^{ \mu} s_{ \Delta}  =   \Delta (  \Delta -4   ) s_{ \Delta}.   \label{eom} \ee
We use Greek indices $ \mu, \nu$ for $AdS_5$ , and $ \alpha, \beta $ for $T^{1,1}$. $Y_n( \psi, \nu_i )$ is  a  function  given in \eqref{sph_h}. The relevant aspects of the spherical harmonics of $T^{1,1}$ are given in appendix \ref{harmonics}. The conformal dimension $ \Delta$ is related with a positive integer $n$  by $\Delta=3n/2$.

  $ c( \Delta)$ is chosen to normalize the two point function of chiral operators as eq.\eqref{2pt}. $ c( \Delta)$ is obtained following \cite{Freedman:1998tz, Lee:1998bxa} , as reviewed in appendix \ref{normalization}, 
\be c( \Delta ) =  \frac{n+1}{ C_n 3^{3/2} N } \sqrt{ \frac{ \Delta +1}{ \Delta ( \Delta +2) } } \label{c-normalization}. 
 \ee

The linearized fluctuation of the  D3-brane action is 
\begin{align}
& { \cal S }^{(1)}_{ DBI} = \frac{ T_{D3} }{2} \int  d^4 \xi \sqrt{  \det G} G^{ mn} \left( \del_m X^{ \mu} \del_n X^{ \nu} h_{ \mu \nu}^{ AdS} + \del_{ m}  X^{ \alpha} \del_n X^{ \beta} h_{ \alpha \beta}^{T^{1,1} } \right) , \nn \\
&  { \cal S }^{(1)}_{ WZ} = T_{D3} \int a^{AdS} . \nn 
\end{align} 
where $T_{ D3}$ is the tension of the D3-brane and fluctuation of fields $h_{ \mu \nu}, h_{ \alpha \beta}, a^{AdS}$ are as defined in \cite{Drukker:2008wr, Kim:1985ez}. $G_{ mn}$ is the induced metric on the world-volume. 
We choose the following world-volume coordinates 
\be \xi^{ m} = \{ x^1, x^2, x^3, x^4 \} \nn \ee
where $x^4$ is Wick-rotation of $x^0$ from \eqref{world}, since we are now considering Euclidean $AdS$. 
The transverse directions are as given in \eqref{transverse}. 
The linearized DBI action for the ansatz becomes, 
\ba { \cal S }_{ DBI }^{(1)} &=&  \frac{ T_{ D3} }{2}   \int d^4 \xi \frac{1 }{ \kappa^2 y^4} \left( \kappa^2  y^2 ( h_{11}^{ AdS} + h_{22}^{ AdS} ) + ( 1 + \kappa^2)  y^2  ( h_{33}^{ AdS} + h_{44}^{ AdS} ) + y^2  h_{yy}^{ AdS} + 9    h_{ \psi \psi}^{T^{1,1} } \right)   
\nn  \\
&& +   \frac{ T_{ D3} }{2}    \int d^4 \xi \frac{1 }{ \kappa^2  y^4}   \left( 2y x^1  h_{ 1 y}^{ AdS} + 2y x^2 h_{ 2 y}^{ AdS} \right) . 
\nn 
\ea
The relations between the source and the fluctuation are given in \cite{Kim:1985ez, Lee:1998bxa, Drukker:2008wr},   
\begin{align} h_{ \mu \nu}^{ AdS} &= - \frac{6}{5} \Delta s g_{ \mu \nu} + \frac{4}{ \Delta +1} \nabla_{ ( \mu } \nabla_{ \nu) } s ,   \nn   \\
a_{ \mu \nu \rho \sigma}^{ AdS} &= - 4 \sqrt{ g^{ AdS} } \epsilon_{ \mu \nu \rho \sigma \eta } \nabla^{ \eta} s  \nn  \\
h_{ \alpha \beta}^{ T^{1,1} } &= 2  g_{ \alpha \beta}  \Delta s  \nn 
\end{align}
where $g_{ \mu \nu}$($ g_{ \alpha \beta}$) is the space-time metric of $AdS_5$ ($T^{1,1}$). $ \nabla_{( \mu} \nabla_{ \nu)} $ is symmetric and trace-less part of $ \nabla_{ \mu} \nabla_{ \nu} $. 

Let  the position of the source be $ x^{\prime} = (0, x^{1 \prime} , x^{2 \prime} , 0 ) $. Indicate the polar coordinates of the source as $ (d, \phi_0)$,   $ d e^{ i \phi_0} \equiv  x^{ 1 \prime} + i x^{ 2 \prime}   $. If we denote numerator of the propagator $ L^{  \Delta}$, 
\be L (y, x: x^{ \prime} ) \equiv y^2 + \sum_{i=0}^3 ( x^i - x^{i \prime } )^2,  \nn \ee
then the linearized fluctuations of DBI and WZ action are written as follows, 
\begin{align} { \cal S }_{ DBI}^{(1)} 
&= 4 \Delta T_{ D3}  \int d^4 \xi \frac{1}{y^4} \left( -1 + 2 \frac{ y ( d / \kappa ) \cos ( \phi - \phi_0 ) }{ L}  - 2  \frac{ y^2  (d/ \kappa)^2  }{L^2}  \right) s   , 
\nn \\
{ \cal S }_{ WZ}^{(1)} 
&= 4 T_{ D3} \Delta \int d^4 \xi \frac{1}{y^4} \left( - 1 + 2 \frac{y ( d / \kappa ) \cos( \phi - \phi_0 ) }{ L}  \right) s . 
\nn 
\end{align} 
Since the action of the D3-brane is $ { \cal S} = { \cal S }_{ DBI}- { \cal S }_{ WZ} $,  the linearized fluctuation of the D3-brane action becomes
\begin{align} 
{ \cal S}^{(1)} &   = - 8 \Delta T_{ D3} \int d^4 \xi \frac{s}{  L^2 } 
\label{L_1} .
\end{align}
We now take the derivative of ${ \cal S}^{(1)}$ with respect to $ s_0^{ \Delta} $ to get the one point function of $ { \cal O}_{ \Delta}$, 
\begin{align}
  \frac{ \langle { \cal O}_{ \Sigma} \cdot { \cal O}_{ \Delta} \rangle }{ \langle { \cal O}_{ \Sigma} \rangle } = - 
  \frac{ \delta { \cal S}^{(1)} }{ \delta s^{ \Delta  }_0 }  & =  8 \pi  T_{ D3}  \frac{ \Delta}{ \Delta +1} c ( \Delta ) C_{ n } \left( \frac{d}{ \kappa} \right)^2  \int dx^1 dx^2 \frac{ y^{ \Delta- 2} (r)  e^{ i \frac{ n }{2} \psi ( \phi)  } }{ ( y^2 (r)  + \vert z_1 - d e^{ i \phi_0}  \vert^2 )^{ \Delta +1} } .
 \label{result:1} 
 \end{align}  

A caveat here is that there are two values of $\psi$ in $0\le \psi \le 4\pi$ for one value of $\phi$ in $0\le \phi \le 2\pi$ as seen in \eqref{transverse}
\be \psi = - 3\phi,\quad -3\phi  + 2 \pi.  \label{2nd} \ee
The result is the sum of contributions from the both branches.  The second branch contributes as much as the result in the first branch multiplied by $(-1)^n$. Thus the result vanish when $n$ is an odd integer.

The integral \eqref{result:1} can be evaluated exactly as in \cite{Drukker:2008wr}.  The result \eqref{result:1} is non-trivial  when $ \Delta  $ is an integer, and it is  given by
\begin{equation}
\begin{aligned}
 \frac{ \langle { \cal O}_{ \Sigma} \cdot { \cal O}_{ \Delta} \rangle }{ \langle { \cal O}_{ \Sigma} \rangle }  
&= 16 \pi^2 T_{ D3} \frac{ c( \Delta) C_n }{   \Delta +1  } \left( \frac{ \kappa}{d e^{ i \phi_0 } } \right)^{ \Delta } , \qquad (\mbox{for} \quad  n \in 2 { \Zb}_{+} )\\
&=\frac{\sqrt{3}}{2}\frac{2\Delta+3}{\sqrt{\Delta(\Delta+1)(\Delta+2)}}  \left( \frac{ \kappa}{d e^{ i \phi_0 } } \right)^{ \Delta } ,
\end{aligned} \label{result:2}  
\end{equation}
where we used $T_{D3}=\frac{3^3N}{2^5\pi^2}$ in the unit of eq.\ \eqref{unit}, and the value of $c(\Delta)$ in \eqref{c-normalization}.

Let us assume that \eqref{gauge} is valid, and compare the result in \eqref{result:2} with the result in the gauge theory \eqref{gauge}.
Firstly, spacetime dependences of those two results agree to each other though they are determined completely by the symmetry.
Secondly, since both $\beta$ and $\kappa$ are parameters of the surface operator and independent of $\Delta$, we can conclude that they are related by
\begin{align}
 \kappa=\mu^{-1/3}\beta^{4/3} \label{id},
\end{align}
with a constant $\mu$. This $\mu$ can only depend on the parameters of the theory.
Thirdly, by comparing the coefficients, we obtain the relation
\begin{align}
 p_{\Delta}C_n=\frac{\sqrt{3}}{4}\frac{2\Delta+3}{\sqrt{\Delta(\Delta+1)(\Delta+2)}}\mu^{-n/2}.
\end{align}

\section{Discussion} \label{discussion} 

In this paper, we studied 1/2 BPS surface operators in the Klebanov-Witten theory. We defined the surface operator by imposing the boundary condition at a codimension 2 singularity. This boundary condition comes from a 1/2 BPS configuration of the bifundamental scalar fields \eqref{example}. We also proposed the gravity dual of this surface operator as a D3-brane configuration \eqref{transverse}. We checked the supersymmetry of this D3-brane. It turned out that this D3-brane preserves half of the supersymmetries of $AdS_5\times T^{1,1}$ , which is consistent with the supersymmetry of the surface operator. We also calculated the correlation function of the surface operator and a local operator. We compared this result with the classical approximation in the gauge theory result and found the qualitative agreement and some constraint on the normalization constants.

For more quantitative matching between the gauge theory side and the gravity side, one needs to calculate the quantities in the gauge theory side. Those techniques like chiral ring and Konishi anomaly equation in \cite{Cachazo:2002ry} could be useful for this purpose.

One of possible interesting future works is to consider the surface operators in other ${\cal N}=1$ superconformal field theories. In particular the quiver gauge theories derived from toric Sasaki-Einstein manifolds \cite{Franco:2005rj,Franco:2005sm,Butti:2005sw}(see also \cite{Yamazaki:2008bt} for a review) will be interesting from the point of view of the AdS/CFT correspondence.

Let us briefly explain a possible extension of the surface operator to other quiver gauge theories.
First we consider a half BPS surface operator with the surface supported at $ z_1 = 0$. We define a surface operator by 
the following boundary conditions for a scalar $ \varphi$ in the chiral superfield, which is in the bi-fundamental representation of $G_I \times G_J$, and the gauge fields $A$ and $ \tilde{A}$ of $G_I$ and $G_J$,  
\begin{align}
& \varphi = \frac{1}{z_1^{( 1 + \gamma_{ \varphi}  ) } } \beta_{ IJ}  , \quad A = \alpha_I  d \phi, \quad \tilde{A} =  \alpha_J d \phi.   \label{def_n1} 
\end{align}
$ \gamma_{ \varphi} $ is the anomalous dimension of the scalar field $ \varphi$ at the IR fixed point.  $ \beta, \alpha, \tilde{ \alpha}$ are constant matrices satisfying 
\be
  \alpha_I \beta_{ IJ} - \beta_{IJ}  \tilde{\alpha}_J  = 0 . \label{commutator}
\ee
 for each $I, J$. If one can find $\beta_{ IJ} , \alpha_I $'s such that  all the $F$ and $D$ term vanish, then by imposing the condition \eqref{half}, the variation of fermions \eqref{quark}, \eqref{gaugino} vanish. Therefore it defines a half BPS surface operator. $1/8$ BPS surface operators of $\N=4$ SYM is a special case for this \cite{Koh:2008kt}, preserving 2 supercharges. If $\gamma_{\varphi}$'s are rational numbers, the possible monodromy in \eqref{def_n1} can be canceled by an appropriated gauge transformation and this configuration and the surface operator is well-defined. If $\gamma_{\varphi}$'s are irrational numbers, then there is no surface operator of this kind.

Second we can also consider the case that $ \varphi$ is a holomorphic function of $z_1$ and $z_2$, 
\be \varphi = \frac{1}{ f( z_1, z_2 )  } \beta_{IJ} ,   \nn \ee
where $f( z_1, z_2)$ is a locally holomorphic homogeneous function with degree $1+ \gamma_{ \phi} $. The surfaces are now supported at $ f( z_1, z_2) = 0 $.  It also preserves $1/2$ of the supersymmetries, since \eqref{half} implies 
\be ( \sigma^3 + i \sigma^4 ) \epsilon = 0,  \label{z2} \ee
for $ \sigma^4 = i I_2$.

The gravity dual of these surface operators would be the D3-brane configuration in $AdS_5 \times SE_5$ considered in section \ref{susy_brane}. If the periodicity of $\zeta\psi$ is (rational number)$\times 2\pi$, this probe D3-brane is closed and corresponds to the surface operator. If the periodicity of $\zeta\psi$ is (irrational number)$\times 2\pi$, or $\zeta\psi$ is not periodic, then this D3-brane cannot be closed. This unclosed D3-brane seems to correspond to the failure of the definition of the surface operator when $\gamma_{\varphi}$ is an irrational number.

\subsection*{Acknowledgments}
We would like to thank Nadav Drukker, Tohru Eguchi, Dongmin Gang, Sangmin Lee, Sanefumi Moriyama, Koichi Murakami, Muneto Nitta, Soo-Jong Rey,  Katsuyuki Sugiyama, and Takao Suyama for useful discussions.  SY is also grateful to the organizers and the participants of ``Taiwan String Theory Workshop 2009,'' at National Taiwan University (Jan.\ 19-21, 2009) and ``KEK Theory Workshop 2009'' at KEK, Tsukuba (Mar.\ 16-19, 2009) for the useful discussions and the hospitality. 
This work was supported in part by KOFST BP Korea Program, KRF-2005-084-C00003, EU FP6 Marie Curie Research and Training Networks MRTN-CT-2004-512194 and HPRN-CT-2006-035863 through MOST/KICOS.

\appendix

\section{Killing spinor of $AdS_5 \times SE_5$}  \label{spinor}

Killing spinor of $AdS_5 \times SE_5$, which makes the variation of the gravitino in \eqref{gravitino} vanish, is given in 
\cite{Kehagias:1998gn} (see also \cite{Yamaguchi:2003ay}). Let the vielbeins of $AdS_5 \times SE_5$  be 
\begin{align} &  E^i = y^{-1} dx^{ i }   \quad (  i = 0, 1, 2, 3) , \quad  E^4 = y^{-1} dy,  \nn \\
&  ds^2_{ SE_5} = \sum_{ a = 5}^9 E^a \otimes E^a , \label{vie:se5} 
\end{align} 
we use the following representation of gamma matrices 
\begin{align}
&  \Gamma^{ \mu } = \tilde{ \Gamma}^{ \mu}  \otimes I \otimes \sigma_1 , \quad ( \mu = 0, \ldots, 3) , \quad  \Gamma^4 = \tilde{ \Gamma}^4 \otimes I \otimes \sigma_1,  \nn \\
& \Gamma^a = I \otimes \tilde{ \Gamma}^a \otimes \sigma_2, \quad ( a = 5, \ldots, 9 )  \label{tilde}
\end{align}
where $\tilde{ \Gamma}^{ \mu} , \tilde{ \Gamma}^4$ ( $ \tilde{ \Gamma}^a $ ) satisfy $SO(1,4)$ ( $SO(5)$) Clifford algebra, and $ \sigma_i$ are Pauli matrices. We choose $ \tilde{ \Gamma}^A$ to satisfy the followings, 
\be \tilde{ \Gamma}^4 = i \tilde{ \Gamma}^{0123}, \quad \tilde{ \Gamma}^{56789} = 1 . \nn \ee

The Killing spinor is in the form of 
\be \epsilon_{ \pm} = \lambda_{ \pm} \otimes \chi \otimes \begin{pmatrix} 1 \\ 0 \end{pmatrix} \label{decomposed} \ee
where $ \lambda_{ \pm}, \chi$ are 4-component spinors. $ \lambda_{ \pm}$ is given in terms of constant spinors $ \eta_{ \pm} $ satisfying $ i \tilde{ \Gamma}^{0123} \eta_{ \pm} = \pm \eta_{ \pm} $, 
\be
\begin{cases} \lambda_{+} &= y^{- \half} ( x_{ \mu} \tilde{ \Gamma}^{ \mu 4 } \eta_{-} + \eta_{+} ) = y^{ - \half} \left( - x_{ \mu} \tilde{ \Gamma}^{ \mu} \eta_{-} + \eta_{+} \right),   \\
\lambda_{-} & = y^{ \half} \eta_{-} . 
\end{cases}
\label{lambda}
\ee
$ \chi$ satisfies the following equation, 
\be \left( \del_a + \frac{1}{4} \omega_a^{ \phantom{a} bc} \tilde{ \Gamma}_{ bc} \right) \chi + \frac{i}{2} \tilde{ \Gamma}_a \chi = 0.  \quad ( a = 5, 6, \cdots, 9)  \label{k:se5} \ee

\section{Proof of  eq.\ (3.18)} \label{proof}

 Let the vielbein of a toric $CY_3$, which is the cone of $SE_5$, be $ \Theta^a$, 
\be ds^2 = dR^2 + R^2 ds^2_{ SE_5} =  \sum_{a=1}^{ 6} \Theta^a \otimes \Theta^a    \nn \ee
where 
\be 
 \Theta^a = R E^{ a+4} \quad ( a = 1, 2, \cdots, 5) , \quad \Theta^6 = dR.  \nn \ee
for given vielbeins $E^a$ of $SE_5$ in \eqref{vie:se5}. 
Define $SO(6)$ gamma matrices $\hat{ \Gamma}^i $ as follows
\be  \hat{ \Gamma}^a = \tilde{ \Gamma}^{ a+4} \otimes \sigma^1, \quad \hat{ \Gamma}^{ 6} = I_4  \otimes \sigma^2 , \quad ( 1 = 1, \cdots, 5) \label{g:cy3} \ee
where $ \tilde{ \Gamma}^a$ are given in \eqref{tilde}. 
Consider a spinor $ \hat{ \chi}$  satisfying 
\be \nabla_i \hat{ \chi} = 0 .  \quad ( i = 1, \cdots, 6) \label{k:cy3} \ee
Then a 4-component spinor $\chi$ such that 
\be \hat{ \chi} = \chi \otimes \begin{pmatrix} 0 \\ 1 \end{pmatrix} , \label{chi} \ee
satisfies \eqref{k:se5} and $ \del_R  \chi = 0 $. 

We choose a local frame  to set the holomorphic 3-form $ \Omega$ and K\"{a}hler 2-form $J$ in the following form 
\be \Omega = ( \Theta^1 + i \Theta^6)( \Theta^2 + i \Theta^3) ( \Theta^4 +i \Theta^5), \quad J = \Theta^1 \Theta^6 + \sum_{ i = 1}^2 \Theta^{ 2i} \Theta^{ 2 i +1} . \nn \ee
In this frame, the Killing spinor $ \hat{ \chi}$ in \eqref{chi} is given by constant spinors $ \hat{ \chi}_{ \pm}$, defined by 
\begin{align} &  \left( \hat{ \Gamma}^{ 1} + i \hat{ \Gamma}^6 \right) \hat{ \chi}_{-} = \left(    \hat{ \Gamma}^2 + i \hat{ \Gamma}^3  \right) \hat{ \chi}_{-} =  \left(  \hat{ \Gamma}^4 + i \hat{ \Gamma}^5  \right) \hat{ \chi}_{-} = 0,    \nn  \\
& \hat{ \chi}_{+} = ( \hat{ \Gamma}^1 - i \hat{ \Gamma}^6 ) ( \hat{ \Gamma}^2 - i \hat{ \Gamma}^3 ) ( \hat{ \Gamma}^4 - i \hat{ \Gamma}^5 )  \hat{ \chi}_{-} . 
\nn 
\end{align}
 By the construction \eqref{g:cy3}, $   \hat{ \Gamma}^{ 123456  } =  i  I \otimes \sigma_3 $. Then the  $ \hat{ \chi}_{-} $ can be  consistent with \eqref{chi}, which implies that 
\be \left( 1 + \tilde{ \Gamma}^5 \right) \chi = \left( \tilde{ \Gamma}^6 + i \tilde{ \Gamma}^7 \right) \chi = \left( \tilde{ \Gamma}^8 + i \tilde{ \Gamma}^9 \right) \chi  = 0.  \label{ks:se5}  \ee

\section{Spherical Harmonics in $T^{1,1}$}  \label{harmonics}

Spherical harmonics $T^{ p,q}$ have been worked out in \cite{Gubser:1998vd}.
In this paper, we only need the spherical harmonics corresponding to the chiral primary operators. They are parametrized by three numbers $(k,m_1,m_2)$ which satisfy
\begin{align}
 k\in \Zb/2,\qquad |m_j|-|k|\in \Zb_{\ge 0}.\label{km}
\end{align}
The spherical harmonics of this kind can be expressed as
\be Y_{ m_1, m_2, k } = C_{ m_1, m_2, k} \left( \cos \frac{ \theta_1}{2} \right)^{ \vert k + m_1 \vert } \left( \sin \frac{ \theta_1}{2} \right)^{ \vert k- m_1 \vert } \left( \cos \frac{ \theta_2}{2} \right)^{  \vert k+ m_2 \vert } \left( \sin \frac{ \theta_2}{2} \right)^{ \vert k - m_2 \vert }  e^{ i ( m_1 \nu_1 + m_2 \nu_2 + k \psi ) }. \label{mmk}  \ee

We use the parameterization \cite{Candelas:1989js} of $ \omega^1, \omega^2, \omega^3, \omega^4$  as follows, 
\be
\begin{aligned}
& \frac{1}{ \sqrt{2}} (   \omega^1 +i \omega^4 ) = y^{-3/2} e^{ \frac{i}{2} ( \psi - \nu_1 - \nu_2 ) } \sin \frac{ \theta_1}{2} \sin \frac{ \theta_2}{2}  , \quad  \frac{1 }{ \sqrt{2} } (  \omega^1  - i \omega^4 ) = y^{-3/2} e^{ \frac{i}{2} ( \psi + \nu_1 + \nu_2   ) } \cos \frac{ \theta_1}{2} \cos \frac{ \theta_2}{2} ,  \\
&  \frac{1}{ \sqrt{2} } ( i  \omega^2 + \omega^3 ) = y^{-3/2} e^{ \frac{i}{2} ( \psi + \nu_1 - \nu_2 ) }  \cos \frac{ \theta_1}{2} \sin \frac{ \theta_2}{2}, \quad   \frac{1}{ \sqrt{2} }   ( i \omega^2 - \omega^3 )  = y^{-3/2} e^{ \frac{i}{2} ( \psi -  \nu_1 + \nu_2  ) } \sin \frac{ \theta_1}{2} \cos \frac{ \theta_2}{2}  . 
\end{aligned}
\label{parameter}
\ee

On the curve $ \omega^2 = \omega^3  = 0 $,  $ \omega^1 = i \omega^4 $, or   $ \theta_i = \pi $ as in \eqref{transverse}, 
 the spherical harmonics is non-zero only when 
\be m_1=  m_2 =  -  k  .\nn \ee

From \eqref{km}, $k$ can be written in terms of an integer $n$,  $ k = \frac{n}{2}$. Finally, we can rewrite the spherical harmonics in our interest as follows 
\be Y_n ( \psi, \nu_1, \nu_2 ) = C_n e^{ i \frac{ n}{2} ( \psi - \nu_1 - \nu_2 )  } , 
\label{sph_h} \ee
where $ C_n = C_{ -\frac{n}{2}, - \frac{n}{2}, \frac{n}{2}  } $ in terms of 
 $C_{ m_1, m_2, k}$ defined in \eqref{mmk}. 
 
 \section{Normalization of $c( \Delta)$}  \label{normalization} 
In this appendix, we explain how to determine the normalization constant $c(\Delta)$ in eq.\ \eqref{c-normalization}.

Let us consider the bulk action of a complex scalar $s$ in $AdS_5$ , 
\be S (s)   = \eta ( \Delta ) \int_{ AdS_5}  d^5 x dy \sqrt{ g}  \left( \vert \nabla s \vert^2 + m^2 |s|^2 \right), \nn \ee
for $m^2 = \Delta ( \Delta -4)$. For the source \eqref{source},  the correct two point function of chiral operators  can be found in  \cite{Freedman:1998tz}, where a subtle change from GKPW prescription has been fixed by bootstrap methods; the two point function and three point functions are related via Ward identity. The result is given by 
\be \frac{ \del^2 }{ \del \bar{s}_0^{ \Delta} (x) \del s_0^{ \Delta } ( x^{ \prime} ) } S =  c^2 ( \Delta )    \eta ( \Delta ) 
\frac{ 2\pi^{ 2} }{\Delta-1} \frac{1}{ \vert x - x^{ \prime} \vert^{ 2 \Delta } } . 
\nn \ee

The normalization of the action $ \eta ( \Delta ) $ is originated from the 10 dimensional IIB supergravity theory, as given in \cite{Lee:1998bxa} (for the detail, see section 3.4 and eq.\ (3.23) of \cite{Lee:1998bxa}),
\be \eta ( \Delta ) = \frac{1}{ 2 \kappa^2} \times 32 \frac{ \Delta ( \Delta -1)( \Delta +2)}{ \Delta +1} \times z( \Delta) , \nn \ee
where $ \frac{1}{ 2 \kappa^2 } = \frac{ 3^6 N^2}{ 2^{11} \pi^5 }$ is the 10 dimensional Newton constant.  $z( \Delta )$ is defined by the normalization of the spherical harmonics of $SE_5$, 
\be \int_{ SE_5} \sqrt{g} Y^{I \ast }_{ \Delta}  Y^J_{ \Delta}  = \delta^{IJ} z ( \Delta ). \nn \ee
For the given spherical harmonics in \eqref{sph_h}, 
\be z ( \Delta ) = \frac{ 2^5 \pi^3 \vert C_n \vert^2 }{ 3^3 ( n+1)^2 }, \nn \ee
where $n$ and $ \Delta$ are related by \eqref{cdim}. 

The form of $c( \Delta )$ in eq.\ \eqref{c-normalization} is determined such that the two point function of chiral operators becomes 
\be \langle \overline{ \cal O}_{ \Delta } ( x)  { \cal O}_{ \Delta }  ( x^{ \prime} ) \rangle = \frac{ 1 }{ \vert x - x^{ \prime} \vert^{ 2 \Delta } } . \nn \ee

\providecommand{\href}[2]{#2}\begingroup\raggedright\endgroup

\end{document}